\documentclass[reprint,amsmath,amssymb,aps,pra]{revtex4-2}
\usepackage{graphicx}% Include figure files
\usepackage{bm}% bold math
\usepackage{tabstackengine}
\usepackage{eucal}
\usepackage{bm}
\usepackage{booktabs}
\usepackage{xcolor}
\usepackage{amsmath, graphics, epsfig, color, verbatim, amsfonts, tensor}
\usepackage{hhline}
\usepackage{multirow}
\usepackage{siunitx}
\usepackage{tabularx}
\usepackage{svg}
\newcommand{\reto}{RE$_2$O$_3$}
\newcommand{\scto}{Sc$_2$O$_3$}
\newcommand{\yto}{Y$_2$O$_3$}
\newcommand{\luto}{Lu$_2$O$_3$}

\begin{document}

%\title{Fast water dissociation on the surface of rare-earth metal oxides from first-principles}
\title{Barrierless Water Dissociation on Rare-Earth Sesquioxide Surfaces from First Principles}

\author{
    Shuxiang Zhou$^1$
}
\email{shuxiang.zhou@inl.gov}
\author{Jay A. LaVerne$^2$}
\author{Hanna Hlushko$^1$}
\affiliation{
    $^1$Idaho National Laboratory, Idaho Falls, ID 83415, USA\\
    $^2$University of Notre Dame, Notre Dame, IN 46556, USA
}

\date{\today}

\begin{abstract}
Water dissociation on metal oxide surfaces is a key elementary step 
in heterogeneous catalysis, photocatalysis, and radiation chemistry, 
yet its mechanistic details on rare-earth (RE) sesquioxides remain 
poorly understood. Here, we investigate water dissociation on the (110) 
surfaces of three cubic bixbyite oxides, Sc$_2$O$_3$, Y$_2$O$_3$, 
and Lu$_2$O$_3$, using molecular dynamics combining \textit{ab initio} calculations with 
on-the-fly machine-learning force field acceleration. By sampling 25 
independent trajectories per material, we obtain an unbiased picture 
of the reaction landscape inaccessible to conventional static 
calculations. Two distinct dissociation pathways are identified: a 
conventional proximal mechanism with a small but finite barrier of 
$\sim$0.1~eV, and a previously unreported distal mechanism that is 
effectively barrierless and energetically preferred at both the 
adsorption and dissociation stages. The low barriers are consistent 
with the periodic array of inherently undercoordinated RE$^{3+}$ 
sites in the bixbyite lattice, suggesting that ordered intrinsic 
coordination defects play a role analogous to stochastic oxygen 
vacancies in conventional oxides. 
\end{abstract}

\maketitle

\section{Introduction}

Water dissociation on metal oxide surfaces is a fundamental process 
underpinning a wide range of scientifically and technologically 
important applications, including heterogeneous catalysis, 
photocatalysis, corrosion, and radiation chemistry~\cite{Thiel1987Interaction,Henderson2002Interaction,Mu2017Structural},
and has been extensively characterized by surface-science methods: temperature-programmed desorption (TPD), high-resolution electron energy loss spectroscopy (HREELS), scanning tunneling microscopy (STM), and X-ray photoelectron spectroscopy (XPS) \cite{Henderson2002Interaction, Brookes2001Imaging, Bikondoa2006Direct, Mu2017Structural}. 
The first O--H bond cleavage, the transfer of a hydrogen atom from an adsorbed water molecule to a surface oxygen site, is often the rate-limiting elementary step that governs the overall 
surface reactivity~\cite{Wang2017Probing}. 
This mechanism has been widely studied computationally on prototypical 
metal oxide surfaces~\cite{Henderson2002Interaction, Mu2017Structural}, 
including MgO~\cite{Leeuw1996Atomistic}, TiO$_2$~\cite{Wang2017Probing, Zeng2023Mechanistic},  
ZnO~\cite{Meyer2006Water}, and UO$_2$~\cite{Bo2014First}. 
The widely accepted mechanistic picture, established through decades of density 
functional theory (DFT) calculations, 
describes a \textit{proximal dissociation} mechanism: the adsorbed 
water molecule coordinates to an undercoordinated surface metal cation 
(Lewis acid site), and the hydrogen atom is transferred to the nearest 
surface oxygen directly bonded to that metal cation, which acts as a 
Brønsted base~\cite{Wang2017Probing, 
Meyer2006Water, Leeuw1996Atomistic}.
On stoichiometric surfaces this process carries a finite 
activation barrier, which varies considerably across systems: 
from 0.13~eV on UO$_2$(111) \cite{Bo2014First} to 0.36~eV on rutile TiO$_2$(110)~\cite{Wang2017Probing}, with the precise 
value being also sensitive to the choice of computational 
approximation~\cite{Zeng2023Mechanistic}.

Oxygen vacancy defects are well known to lower this barrier 
substantially by providing undercoordinated metal sites with enhanced 
Lewis acidity and modified surface 
basicity~\cite{Schaub2001Oxygen, Bikondoa2006Direct, 
Henderson2002Interaction}. However, such vacancies are stochastic 
in nature: their concentration, distribution, and stability depend 
sensitively on preparation conditions and are difficult to engineer 
systematically. This raises a compelling materials design question: 
can the reactivity enhancement associated with oxygen vacancies be 
achieved intrinsically, through a crystal structure that incorporates 
ordered, periodic coordination defects by design? 
The rare-earth sesquioxides \reto\ (RE = Sc, Y, Lu) crystallizing in 
the cubic bixbyite structure (space group $Ia\bar{3}$) offer a 
natural candidate. The bixbyite lattice can be derived from the 
fluorite structure by the ordered removal of one-quarter of the 
oxygen atoms, generating a periodic array of coordinatively 
unsaturated RE$^{3+}$ cation sites that are structurally inherent 
rather than defect-induced~\cite{Lee2020Colossal, Chen2022Water}. 
Despite this structural analogy to reduced fluorite-type oxide 
surfaces, the water 
dissociation mechanism on bixbyite surfaces has never been studied 
computationally, except for a recent study on 
In$_2$O$_3$ that focused on stable structures and surface 
hydrophobicity~\cite{Chen2022Water}.

A key challenge is that the low-symmetry bixbyite surface, 
arising precisely from its ordered vacancy superstructure, presents 
a large number of inequivalent surface adsorption environments. 
This diversity makes comprehensive static DFT sampling impractical: 
through conventional static calculations at pre-selected positions, 
non-intuitive reaction pathways, for example those involving 
non-nearest-neighbor surface sites, cannot be discovered. We 
address this challenge by performing molecular 
dynamics (MD) simulations combining \textit{ab initio} calculations and on-the-fly machine-learning force field (MLFF) 
acceleration~\cite{jinnouchi_fly_2019}, enabling systematic and 
unbiased sampling of the reaction landscape at tractable 
computational cost. In this work, we establish a direct mechanistic 
link between crystal structure and surface reactivity, and suggest a 
broader design principle: oxides with ordered, intrinsic coordination 
unsaturation can sustain fast, barrierless water dissociation without 
relying on stochastic defect engineering.

\section{Methods}

All DFT calculations were performed using the projector augmented-wave (PAW) 
method~\cite{blochl_projector_1994, kresse_ultrasoft_1999} as implemented in the Vienna 
Ab initio Simulation Package (VASP)~\cite{kresse_ab_1993, kresse_efficient_1996}. The 
generalized gradient approximation (GGA) of Perdew, Burke, and Ernzerhof 
(PBE)~\cite{perdew_generalized_1996} was used as the exchange--correlation functional. 
For Sc, Y, O, and H, the \textit{Sc\_sv}, \textit{Y\_sv}, \textit{O}, and \textit{H} PBE pseudopotentials supplied with VASP were used, respectively. 
For Lu, the \textit{Lu\_3} PBE pseudopotential was adopted, treating the $4f$ electrons as 
fully-localized core states. A plane-wave cutoff energy of 500~eV was applied throughout. 
Spin polarization was not included, as Sc$_2$O$_3$, Y$_2$O$_3$, and Lu$_2$O$_3$ are 
nonmagnetic owing to their closed-shell $d^0$ or $4f^{14}$ electronic configurations.
The water dissociation reaction on the surfaces of these oxides was investigated in three 
stages: slab construction, MD reaction simulation, and static energy extraction.

\textbf{Slab Construction.}
Bulk crystal structures were first fully optimized using primitive cells (RE$_{16}$O$_{24}$) with an energy convergence criterion of 
$10^{-7}$~eV/atom and a $\Gamma$-centered $k$-point mesh of 5$\times$5$\times$5. Slab models were then constructed from the 
optimized bulk lattices using the Atomic Simulation Environment  
(ASE) package~\cite{larsen_atomic_2017}. Each slab contains 8 rare-earth oxide layers, sufficient 
to reproduce bulk-like properties in the interior while accurately describing the surface 
region, and is separated from its periodic images by 20~\AA\ of vacuum (10~\AA\ on each 
side). These setups were consistent with similar surface studies in the literature \cite{Bo2014First}. For all slab calculations, energy and force convergence criteria were set to 
$10^{-7}$~eV/atom and $10^{-3}$~eV/\AA, respectively, with only the $\Gamma$ point used for Brillouin zone
sampling.
Slab stability was confirmed by MD simulations, with details provided below. 

\textbf{Reaction Simulation.}
Water adsorption and dissociation were simulated by MD. 
A single H$_2$O molecule was placed 
approximately 10~\AA\ above the slab surface, and assigned an initial 
velocity of 0.02~\AA/fs directed toward the surface. To systematically sample different adsorption 
sites, 25 independent simulations were performed per material, corresponding to a $5 \times 
5$ grid of fractional lateral coordinates $(x, y)$ with $x, y \in \{0, 0.2, 0.4, 0.6, 
0.8\}$. 
All MD calculations were performed in the canonical ($NVT$) ensemble at $T = 300$~K using a 
Nos\'{e}--Hoover thermostat. Each simulation was run 
for 10000 steps with a time step of 0.3~fs, giving a total simulation time of 3~ps, and full atomic trajectories were recorded. In all simulations, water dissociation occurred within this time window.
On-the-fly machine-learning force field \cite{jinnouchi_fly_2019} was enabled 
throughout all MD simulations to improve computational efficiency. 
The MLFF was trained on-the-fly from DFT-computed structures, energies, and forces, and was used for force evaluation whenever the estimated Bayesian error was below a predefined threshold; otherwise, DFT calculations were automatically performed and the resulting configuration was incorporated into the MLFF training set.

\textbf{Energy Extraction.}
All energetic results were computed directly using DFT.
Reaction pathways were identified from the MD trajectories by monitoring (i) the distance 
between the water oxygen (O$_w$) and surface rare-earth atoms, and (ii) the O$_w$--H and 
surface oxygen (O$_s$)--H distances, to assign reaction states (see Supplemental materials [SM] Table~S2). For 
each semistable state persisting for more than 500 consecutive frames (150~fs), one 
representative frame was selected at random from within the central 50\% of the segment 
(i.e., between the 25th and 75th percentile frames), and the structure was fully relaxed at 
0~K to extract the potential energy using DFT.
Furthermore, the energy pathway for the first O--H bond cleavage was determined using the 
climbing-image nudged elastic band (CI-NEB) method~\cite{henkelman_climbing_2000, 
henkelman_improved_2000} as implemented in the VTST code~\cite{henkelman_vtst}. A 
two-stage refinement was employed: an initial set of 5 images was used to map the global 
reaction pathway, after which 5 additional images were inserted in the vicinity of the 
transition state to resolve the energy barrier with greater accuracy.

\section{Results and Discussions}

\textbf{Slab Construction.}
Here, we begin with the bulk crystal properties of \scto, \yto, and \luto; 
the full data, including lattice constants, band gaps, and electronic density of states, 
are reported in the SM Table~S1 and Fig.~S1. The DFT-optimized lattice 
parameters are $a = 9.91$~\AA\ (Sc$_2$O$_3$), $10.669$~\AA\ (Y$_2$O$_3$), and 
$10.35$~\AA\ (Lu$_2$O$_3$), respectively, in good agreements with the respective experimental values 
of 9.87, 10.60, and 10.39~\AA~\cite{Savitskaya1960Formation,Zachariasen1927Crystal,Rykova1982Equilibria}, consistent with the typical GGA 
error of less than 1\%.

The three oxides exhibit strong similarity in their bulk electronic properties: all three 
have lattice constants close to 10~\AA, band gaps of approximately 4~eV, and valence and 
conduction bands dominated by O~$p$ and RE~$d$ orbitals. This similarity 
reflects their common valence electron configuration: Sc, Y, and Lu all adopt a $d^0$ 
electronic configuration in the 3+ oxidation state, resulting in analogous bonding 
character and electronic structure across the three oxides.
Nonetheless, subtle differences always exist. In particular, Lu$^{3+}$ possesses a smaller ionic 
radius than Y$^{3+}$~\cite{Shannon1976Revised}, as a consequence of the lanthanide contraction: the poor 
shielding of the $4f$ electrons leads to a large effective nuclear charge in 
Lu, contracting its electron cloud. This lanthanide contraction effect accounts for the smaller lattice constant of \luto\ compared to \yto.

\begin{figure}[t]
    \centering
    \includegraphics[width=\linewidth]{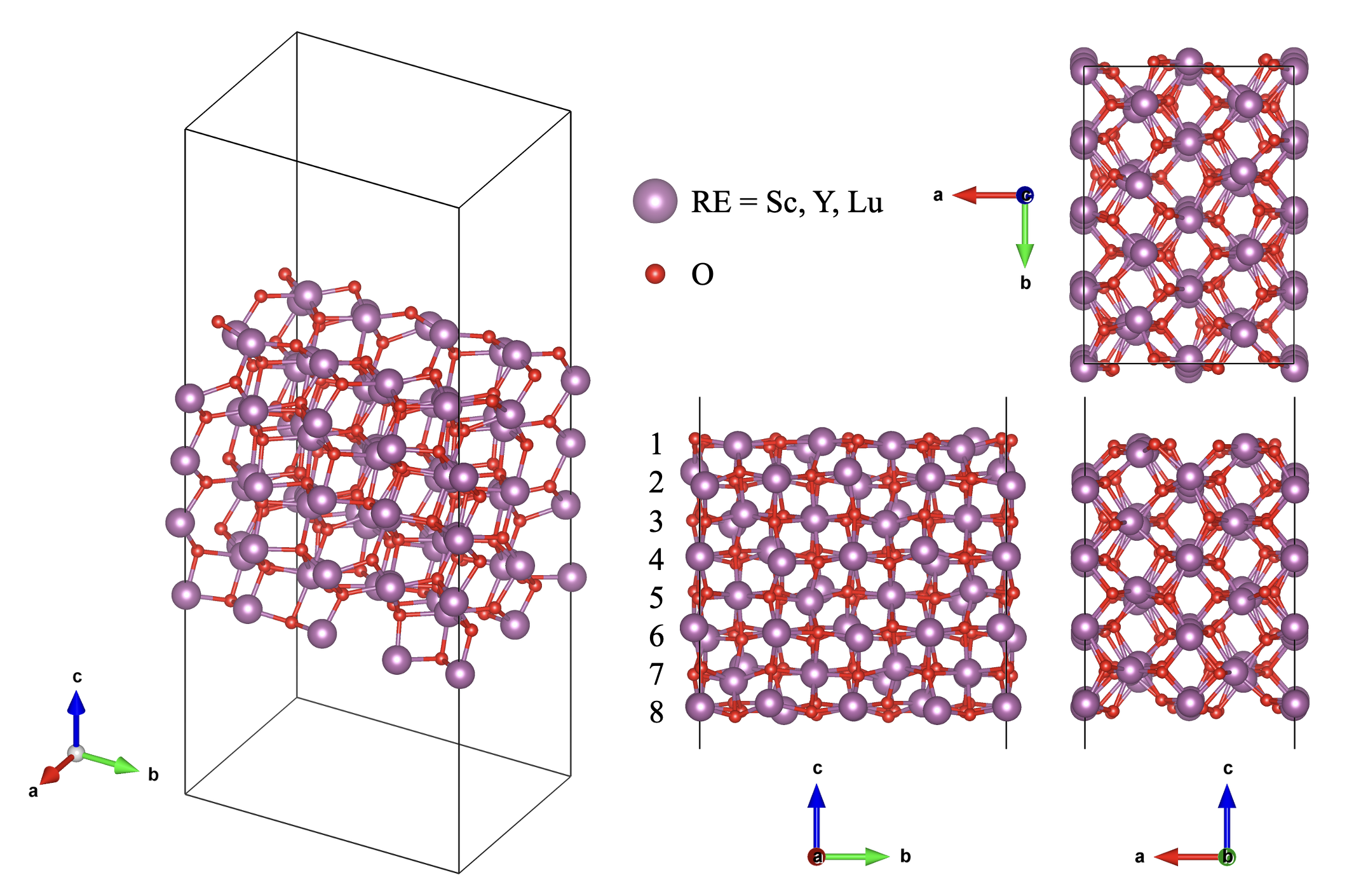} 
    \caption{Slab model of the (110) surface of \reto\ (RE = Sc, Y, Lu). A slab of 8 RE layers is used with a 10~\AA\ vacuum layer applied on both sides. Side views along the three crystallographic axes are also shown.}
    \label{fig:slab_structure}
\end{figure}

Then, slab models were constructed for the (100), (110), and (111) surfaces using the 
DFT-optimized bulk lattice constants. 
%The (111) slab was excluded from further consideration due to its prohibitively large cell of 320 atoms, compared to 160 atoms for both the (100) and (110) slabs. Given that the computational cost of DFT scales as $\mathcal{O}(N^3)$, this twofold increase in system size would incur an approximately eightfold increase in cost, making (111) impractical for the present study. 
The relative stability of the three surfaces was assessed via the surface 
energy,
\begin{equation}
    E_\mathrm{s} = \frac{E_\mathrm{slab} - n E_\mathrm{bulk}}{2A},
    \label{eq:surf_energy}
\end{equation}
where $n$ is the number of formula units in the slab, $E_\mathrm{slab}$ is the DFT 
total energy for the slab, $E_\mathrm{bulk}$ is the DFT 
total energy per formula unit of the bulk crystal, and $A$ is the surface area. 
The computed values of $E_\mathrm{s}$ are reported in the SM Table~S1. 
Across all three oxides, the (110) and (111) surfaces consistently exhibit lower formation 
energies than (100) by approximately 45\%-55\%, indicating that (110) and (111) are thermodynamically more stable.
The preference for (110) and (111) can also be understood from the layer stacking: in the (110) and (111)
orientation, each atomic layer contains both RE and O atoms in the stoichiometric 
ratio, yielding a non-polar, stoichiometric termination. In contrast, the 
(100) orientation consists of alternating pure-O and pure-RE layers, resulting in a 
polar slab whose surface termination depends on where the crystal is cleaved. Such 
polar surfaces are generally less stable and more difficult to model 
unambiguously~\cite{Tasker1979Stability}. 
Comparing (110) and (111), although (111) has a slightly lower formation 
energy, it requires a prohibitively large unit cell of 320 atoms, 
which is twice that of the (100) and (110) slabs (160 atoms each). 
Given that the computational cost of DFT scales as $\mathcal{O}(N^3)$, 
this twofold increase in system size would incur an approximately eightfold increase in cost, 
rendering (111) impractical for the present study. 
Therefore, the (110) surface 
(Fig.~\ref{fig:slab_structure}) was selected for all subsequent simulations. We note, however, that the (111) surface remains experimentally relevant; more broadly, other low-symmetry surfaces with comparably low formation energies may also exist and merit consideration in future studies. The (100) and (111) surfaces are shown for comparison in the SM Fig.~S2.

\begin{figure*}[t]
    \centering
    \includegraphics[width=0.7\linewidth]{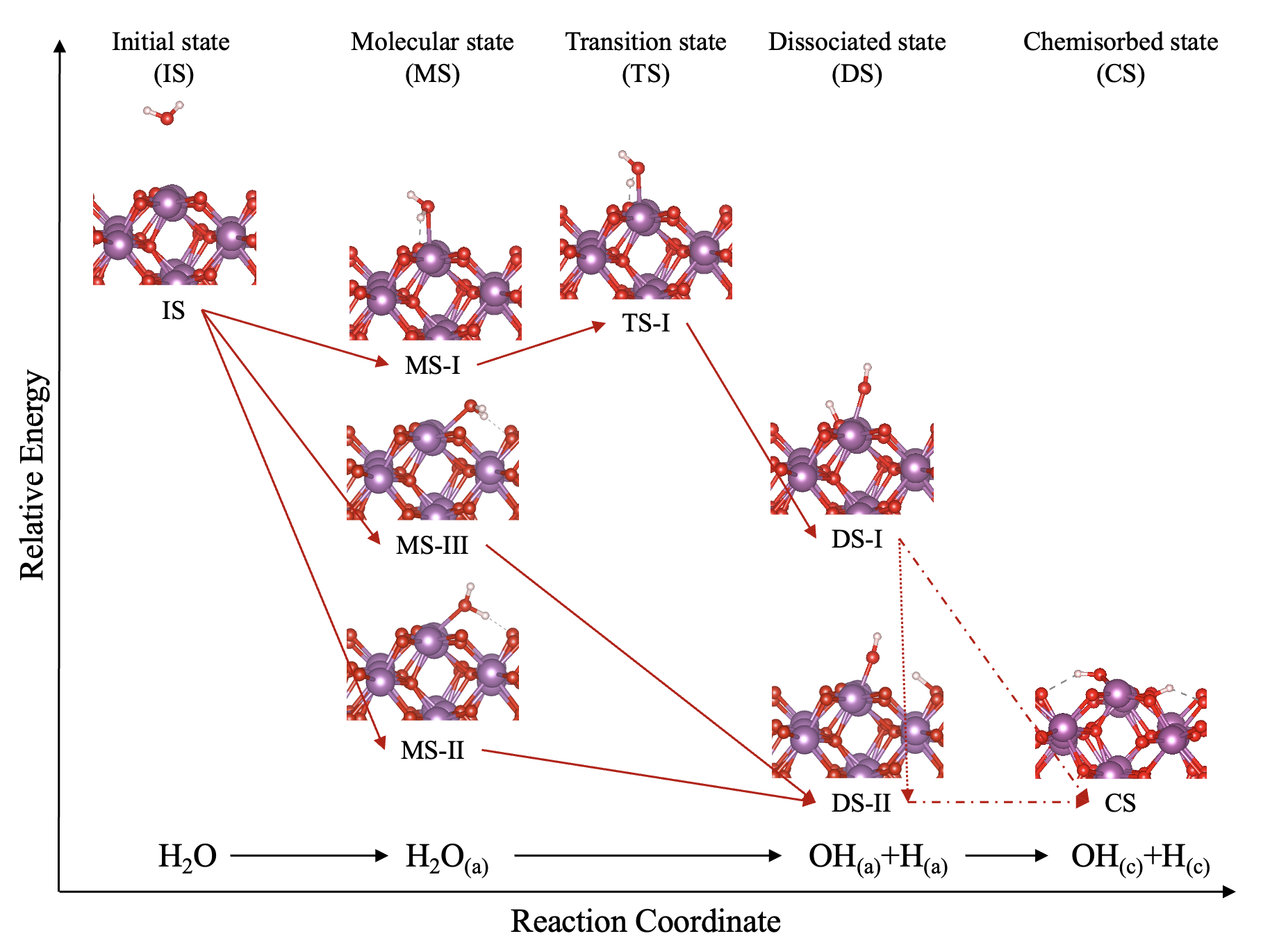}
    \caption{Reaction pathway of water dissociation on the \reto\ (110) surface (RE = Sc, Y, Lu) from MD simulations and CI-NEB calculations. The relative energy is plotted as a function of the reaction coordinate, showing five stages: initial state (IS), molecular state (MS), transition state (TS), dissociated state (DS), and chemisorbed state (CS). Arrows indicate the direction of the reaction pathway. Three molecular adsorption configurations are identified as MS-I, MS-II, and MS-III. MS-I proceeds via TS-I to DS-I, while MS-II and MS-III converge to DS-II without a transition state. For \scto, DS-I further relaxes to DS-II (dotted line); while for \yto\ and \luto, both DS-I and DS-II relax to CS (dashed lines), where OH and H are fully integrated into the surface.}
    \label{fig:pathway}
\end{figure*}

\textbf{Reaction Simulation.}
The reaction states and pathways identified from our MD simulations are summarized in 
Fig.~\ref{fig:pathway}. As discussed above, the water dissociation proceeds through four characteristic stages, referred to hereafter as 
the initial state (IS), the molecular state (MS), the dissociated state (DS), and the 
chemisorbed state (CS). In the IS, both the water molecule and the \reto\ surface remain 
intact. The water molecule subsequently chemisorbs onto a surface RE site, forming the MS (H$_2$O$_{(\text{a})}$). The rate-limiting elementary step follows: one hydrogen atom is 
transferred from the adsorbed water molecule to a surface oxygen site, yielding 
the DS, in which the water molecule has been cleaved into a surface-bound hydroxyl 
(OH$_{(\text{a})}$) and a hydrogen atom (H$_{(\text{a})}$). Finally, the adsorbed OH$_{(\text{a})}$ and H$_{(\text{a})}$ may 
further relax and diffuse into structurally integrated surface configurations, forming the fully hydroxylated 
CS (OH$_{(\text{c})}$ + H$_{(\text{c})}$).

Within this four-stage framework, our MD simulations reveal two distinct dissociation 
pathways, differing in the molecular adsorption geometry and the identity of the hydrogen atom 
acceptor site. In Pathway-I, the water molecule chemisorbs onto a surface 
Lewis acid--base pair (MS-I), and the hydrogen atom is subsequently transferred to a first-nearest-neighbor surface 
oxygen O$_s$ that is directly bonded to the metal cation coordinating the resulting hydroxyl, forming DS-I. Notably, in Pathway-II, the water molecule instead 
chemisorbs with the hydrogen oriented toward a second-nearest-neighbor O$_s$ site that is 
crystallographically non-bonded to the coordinating metal cation, forming either MS-II 
(interaction with one such RE$\cdots$O pair) or MS-III (interaction with two such 
RE$\cdots$O pairs simultaneously). The hydrogen atom is subsequently transferred to this 
second-nearest-neighbor O$_s$, yielding DS-II. 
Therefore, Pathway-I constitutes 
a \textit{proximal dissociation}, consistent with the conventional mechanism 
reported for metal oxide surfaces \cite{Wang2017Probing, 
Meyer2006Water, Leeuw1996Atomistic}. In contrast, pathway-II constitutes a \textit{distal dissociation}, representing a departure from 
the conventional picture. 
Both Pathway-I and Pathway-II are 
directly observed in the MD simulations across all three \reto\ (110) surfaces 
investigated.
Please note that MS-III is rarely observed in the 
MD trajectories, likely due to the geometric constraints imposed by simultaneously 
satisfying two distal H$\cdots$O$_s$ interactions; it is therefore treated as a geometric variant 
of Pathway-II rather than an independent pathway. 

Additionally, the final stable state also differs among the three \reto\ (110) surfaces. For 
\scto, DS-II represents the thermodynamic endpoint with OH$_{(\text{a})}$+H$_{(\text{a})}$. However, for \yto\ and \luto, the endpoint is CS with OH$_{(\text{c})}$+H$_{(\text{c})}$ integrated to the surface. The energetic origins of these differences will be examined in the following section.

\begin{figure}[t]
    \centering
    \includegraphics[width=0.95\linewidth]{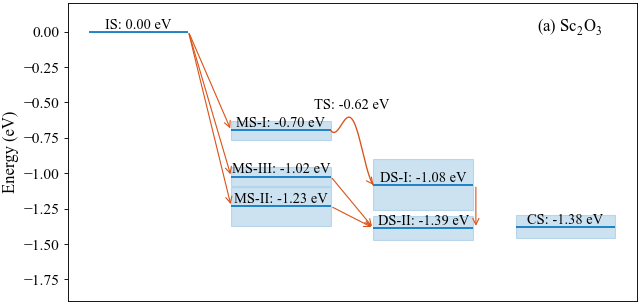}
    \includegraphics[width=0.95\linewidth]{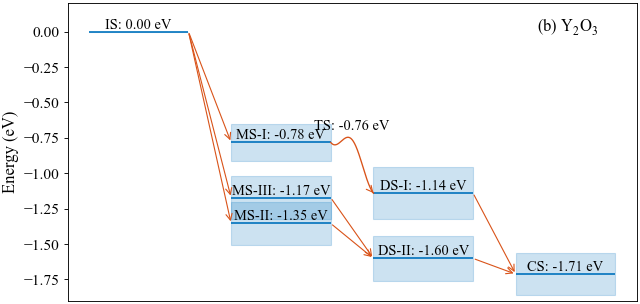}
    \includegraphics[width=0.95\linewidth]{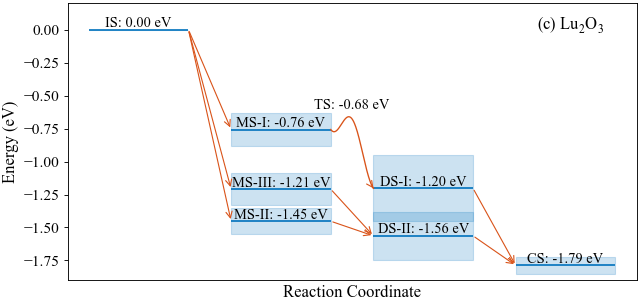}
    \caption{DFT-computed reaction energy profile of water dissociation on the \reto\ (110) surface (RE = Sc, Y, Lu). Energies are referenced to the initial state (IS). The horizontal bars represent the mean energy of each state with shaded regions indicating the standard deviation. The reaction pathway proceeds from molecular adsorption (MS-I, MS-II, MS-III) to the dissociated state (DS) and chemisorbed state (CS). Arrows indicate the direction of the reaction pathway.}
    \label{fig:energy}
\end{figure}

\textbf{Energy Analysis.}
For each state observed in a single MD simulation, one representative frame is randomly 
selected and structurally relaxed to obtain a well-defined potential energy. Only 
relaxed structures that retain the same configuration as the intended state 
are accepted, and the corresponding energies are averaged across all accepted 
frames. The resulting mean energies (horizontal bars), together with their standard deviations (shaded regions), are presented in Fig.~\ref{fig:energy}.
The standard deviation reflects two contributions: numerical errors 
inherent to the DFT calculations, and the structural variability from the 
different adsorption sites due to the low-symmetry \reto\ crystal. Notably, when the 
numerical convergence is well controlled, the energy spread due to site variability is 
of comparable magnitude to the energy release of the water dissociation. This indicates that the choice of adsorption site can significantly influence 
the local reaction energetics on \reto\ surfaces, analogous to the well-known role of 
defect sites in other metal oxide systems.

\begin{figure}[t]
    \centering
    \includegraphics[width=0.9\linewidth]{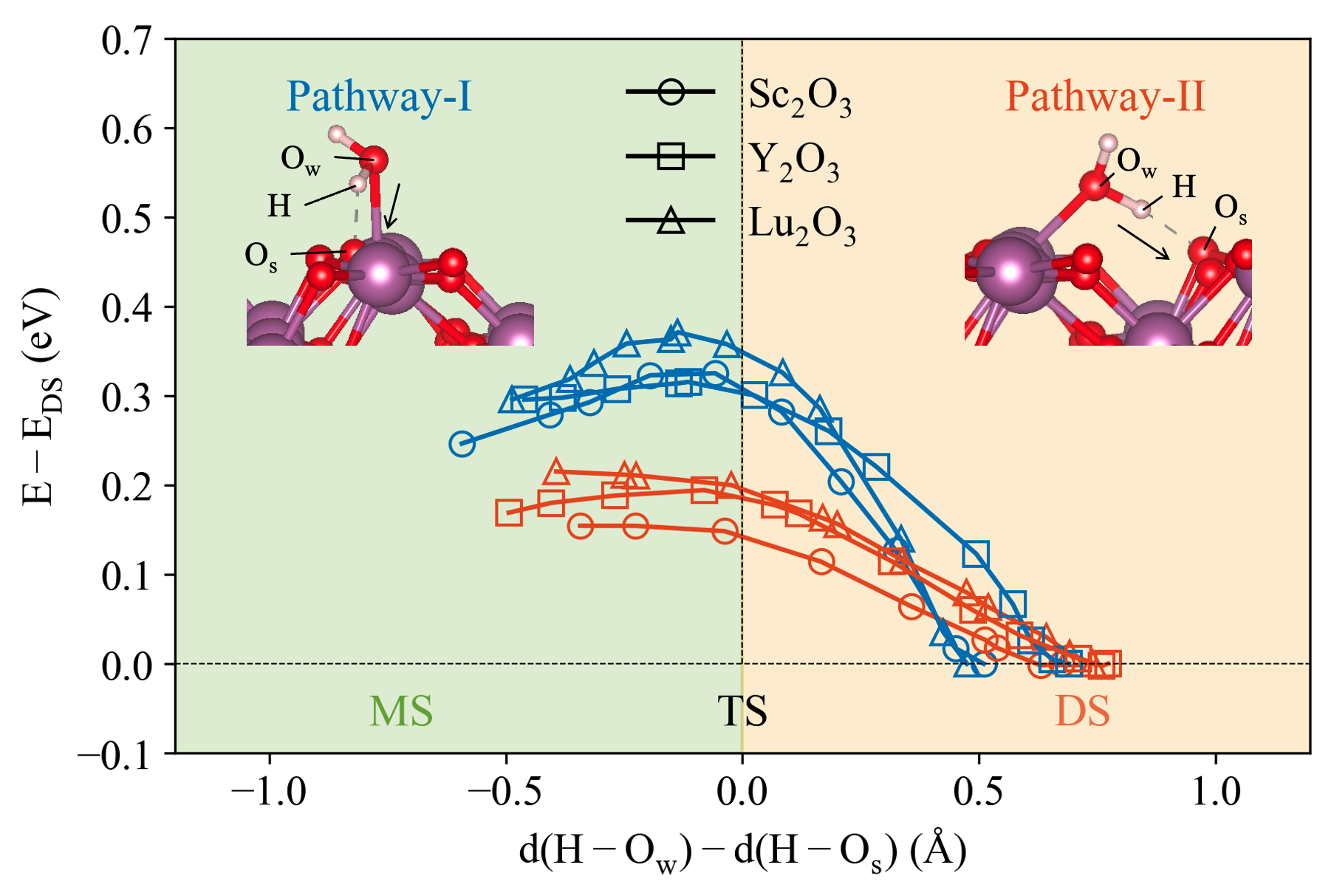}
    \caption{DFT-computed energy along the reaction coordinate $d(\mathrm{H}-\mathrm{O}_w) - d(\mathrm{H}-\mathrm{O}_s)$ for water dissociation on \scto, \yto, and \luto\ (110) surfaces. A hydrogen atom from an adsorbed H$_2$O molecule (with oxygen denoted O$_w$) is transferred to a surface oxygen (denoted O$_s$), where $d(\mathrm{H}-\mathrm{O})$ denotes the distance between H and O. The energy is referenced to the dissociated state (DS). Curve color distinguishes the two pathways (see main text for details), while marker shape denotes the material.}
    \label{fig:transition}
\end{figure}

Several key trends are evident from Fig.~\ref{fig:energy}. Comparing the total energy 
release across the three \reto\ compounds, \yto\ and \luto\ yield closely similar results, 
while \scto\ releases slightly less energy overall; nevertheless, all three follow the 
same qualitative trends, consistent with their similar chemical nature. Across all three 
compounds, MS-II and MS-III are substantially more stable than MS-I, which explains why 
MS-II is frequently observed in the MD trajectories 
despite the geometric constraint of bridging a distal H$\cdots$O$_s$ pair to a second-nearest-neighbor 
O$_s$. Similarly, DS-II is more stable than DS-I in all three cases, 
collectively confirming that Pathway-II is energetically preferred over the conventional 
Pathway-I at both the MS and DS stages.

Since the first O--H cleavage (MS$\rightarrow$DS) is the rate-limiting elementary step, the associated 
energy barrier warrants closer examination. For each pathway and each compound, one 
representative MD trajectory is selected in which both the MS and DS relaxations 
retain their intended configurational character, and the minimum energy path is computed 
using the CI-NEB method. The resulting energy profiles are plotted as a function of the 
reaction coordinate $d(\mathrm{H}-\mathrm{O}_w) - d(\mathrm{H}-\mathrm{O}_s)$, where 
$d$ denotes the distance for the H--O pair (Fig.~\ref{fig:transition}). Consistent trends are observed across all three 
compounds. For Pathway-I, a well-defined energy barrier of approximately 0.1~eV is 
identified, with the transition state occurring near $d(\mathrm{H}-\mathrm{O}_w) \approx 
d(\mathrm{H}-\mathrm{O}_s)$. For Pathway-II, the transferring hydrogen atom travels a 
longer distance to reach the distal acceptor oxygen; however, no significant energy 
barrier (i.e., $>$0.03~eV) is observed along the entire pathway. Therefore, relative to 
the conventional proximal Pathway-I, the distal Pathway-II is not only thermodynamically preferred but also 
kinetically more facile.
It should be noted that, in contrast to the state energy calculations where multiple 
structures are sampled and averaged, only a single representative trajectory is used 
for each CI-NEB calculation, which introduces additional uncertainty into the absolute 
barrier values. Nevertheless, the qualitative distinction between the two pathways, 
a finite barrier of approximately 0.1~eV for Pathway-I versus an approximately barrierless profile 
for Pathway-II, is consistently reproduced across all three \reto\ compounds, 
supporting the robustness of this conclusion despite the limited sampling.
The barrier of Pathway-I is smaller than values reported for 
stoichiometric rutile TiO$_2$(110) (0.36~eV~\cite{Wang2017Probing}), consistent with the enhanced reactivity observed on defective oxide 
surfaces where oxygen vacancies create undercoordinated metal 
sites~\cite{Schaub2001Oxygen, Bikondoa2006Direct}. 
%This is consistent with the bixbyite crystal structure of \reto, which can be derived from the fluorite lattice by the ordered removal of one-quarter of the oxygen atoms, thereby generating a periodic array of inherently undercoordinated RE$^{3+}$ sites throughout the bulk structure. 
Unlike stochastic oxygen vacancies in conventional oxides, 
these ordered coordination defects are an intrinsic and uniform 
feature of the bixbyite lattice, suggesting that the low barriers 
observed for both pathways reflect a general characteristic of this 
structural family.
Please note that a systematic symmetry-based classification of the local environments associated with each reaction pathway could provide direct insight into the role of structural disorder in surface reactivity and establish quantitative relationships between vacancy ordering and reaction energetics, making it an important direction for future work.

\begin{figure}[t]
    \centering
    \includegraphics[width=0.9\linewidth]{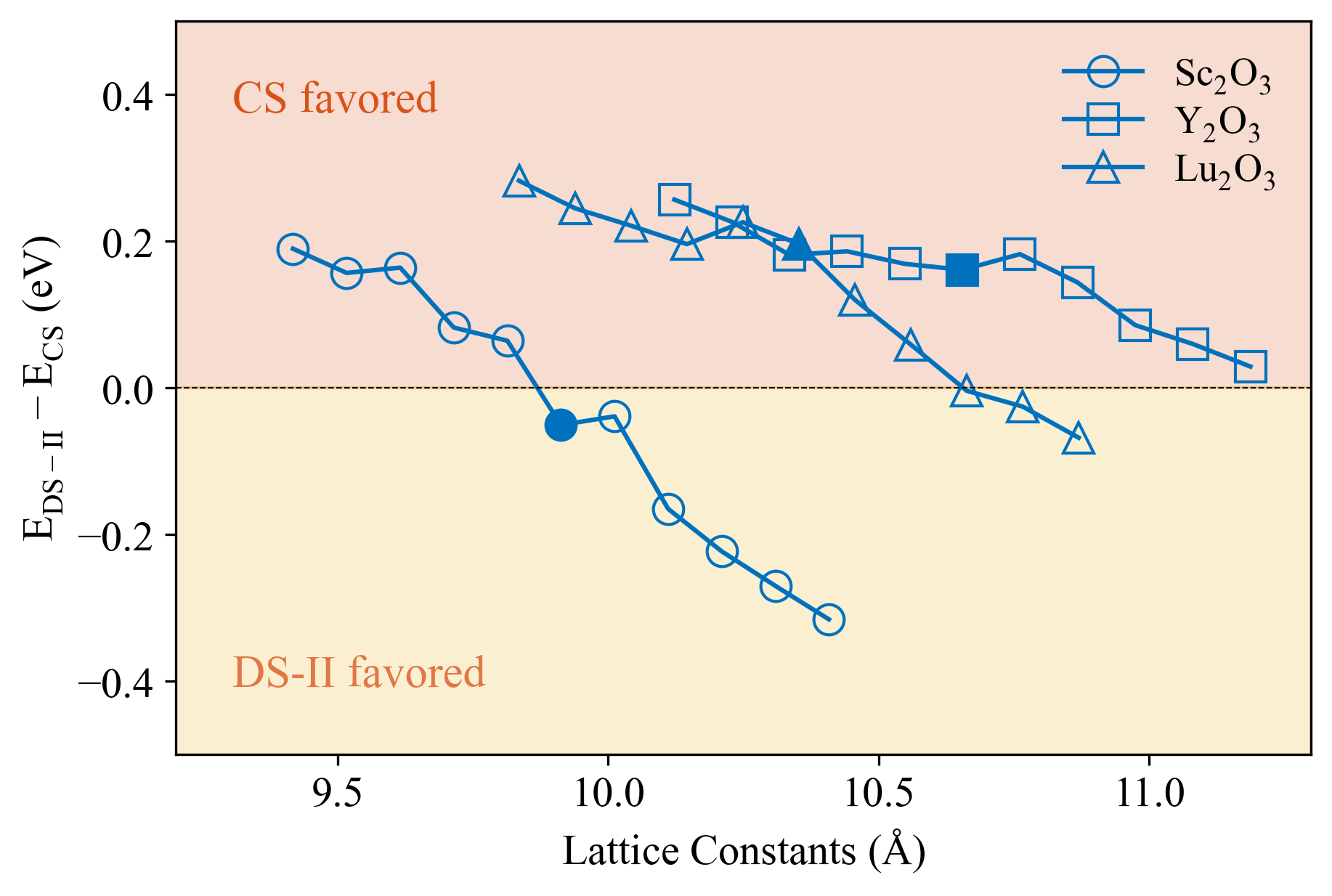}
    \caption{DFT-computed energy difference $E_{\text{DS-II}} - E_{\text{CS}}$ as a function of lattice constant for \scto, \yto, and \luto. Filled markers indicate the equilibrium lattice constant of each material. The yellow and blue shaded regions denote conditions where CS and DS-II are energetically favored, respectively.}
    \label{fig:finalstate}
\end{figure}

Finally, the CS is observed only for \yto\ and \luto, where it represents the global 
energy minimum, indicating stronger stabilization of the fully hydroxylated surface in 
these two compounds. Two factors may contribute to this difference: distinctions in 
chemical bonding character among the three RE cations, and geometric effects 
arising from their different equilibrium lattice constants. To isolate the geometric 
contribution, $E_{\text{DS-II}} - E_{\text{CS}}$ is computed as a function of lattice 
constant for all three \reto\ surfaces (Fig.~\ref{fig:finalstate}). Within each material, 
increasing the lattice constant systematically reduces $E_{\text{DS-II}} - E_{\text{CS}}$, 
making the CS less favorable. However, when comparing across materials at their 
respective equilibrium lattice constants, the trend reverses: larger equilibrium lattice 
constants correlate with greater CS stability. This indicates that the intrinsic chemical 
bonding difference among the rare-earth cations is the dominant factor governing CS 
stability, while the geometric effect of lattice expansion partially offsets this trend. 
Notably, this geometric sensitivity suggests that epitaxial strain or applied pressure 
could serve as a practical handle to tune the thermodynamic endpoint of water 
dissociation between the DS-II and CS configurations, with potential implications for 
controlling the surface hydroxylation state and the adsorption kinetics of subsequent 
water molecules.

\section{Conclusions}

In this work, we have established a mechanistic and energetic picture of water 
dissociation on the (110) surfaces of three cubic bixbyite 
rare-earth sesquioxides, \scto, \yto, and \luto, using \textit{ab 
initio} molecular dynamics with on-the-fly machine-learning force 
field acceleration. By systematically sampling 25 independent 
trajectories per material across a $5\times5$ grid of lateral 
starting positions, we obtained an unbiased picture of the reaction 
landscape that is inaccessible to conventional static DFT 
calculations. 

Two distinct water dissociation pathways were identified across all 
three materials. Pathway-I is the conventional proximal dissociation, 
consistent with the mechanism established for other metal oxide 
surfaces, in which the hydrogen atom transfers to a surface oxygen 
directly bonded to the coordinating RE$^{3+}$ cation within the same 
coordination polyhedron. Pathway-II is a previously unreported 
distal dissociation, in which the hydrogen atom transfers to a 
second-nearest-neighbor surface oxygen that is crystallographically 
non-bonded to the coordinating cation. At both the molecular 
adsorption and dissociation stages, Pathway-II is energetically 
preferred over Pathway-I. CI-NEB calculations confirm that Pathway-I 
proceeds with a small but finite barrier of $\sim$0.1~eV, while 
Pathway-II is effectively barrierless, making it the thermodynamically 
and kinetically dominant pathway. 

The low barriers of both pathways are consistent with the periodic 
array of inherently undercoordinated RE$^{3+}$ sites in the bixbyite 
lattice, suggesting that ordered intrinsic coordination defects may 
play a role analogous to stochastic oxygen vacancies in conventional 
oxides. This principle extends naturally to other members of the rare-earth 
sesquioxide family (i.e., \reto\ with other RE elements) 
and, more broadly, to oxides with intrinsic ordered-vacancy lattices 
beyond this family, providing a rational basis for the computational 
screening and experimental design of oxide catalysts with enhanced 
water activation capability.

\section{Acknowledgments}
This work is supported through the INL Laboratory Directed Research and Development (LDRD) Program under DOE Idaho Operations Office Contract DE-AC07-05ID14517. This research made use of Idaho National Laboratory’s High Performance Computing systems located at the Collaborative Computing Center and supported by the Office of Nuclear Energy of the U.S. Department of Energy and the Nuclear Science User Facilities under Contract No. DE-AC07-05ID14517.

\bibliographystyle{achemso}
\bibliography{reference}

\end{document}

\renewcommand{\thefigure}{S\arabic{figure}}
\renewcommand{\thetable}{S\arabic{table}}
\renewcommand\cellalign{cl}
\newcommand{\reto}{RE$_2$O$_3$}
\newcommand{\scto}{Sc$_2$O$_3$}
\newcommand{\yto}{Y$_2$O$_3$}
\newcommand{\luto}{Lu$_2$O$_3$}

\raggedright{\textbf{\Large Supplemental materials to}}
%\title{Fast water dissociation on the surface of rare-earth metal oxides from first-principles}
\title{Barrierless Water Dissociation on Rare-Earth Sesquioxide Surfaces \\from First Principles}

\author{
    Shuxiang Zhou$^1$, Jay A. LaVerne$^2$, and Hanna Hlushko$^1$
}

\affiliation{
    $^1$Idaho National Laboratory, Idaho Falls, ID 83415, USA\\
    $^2$University of Notre Dame, Notre Dame, IN 46556, USA
}

\maketitle
\setcounter{page}{1}
\renewcommand{\thepage}{S\arabic{page}}
\pagestyle{plain}
\thispagestyle{plain}
%\section{Bulk Calculation results and surface energy}

\begin{figure}[h]
    \centering
    \includegraphics[width=0.51\linewidth]{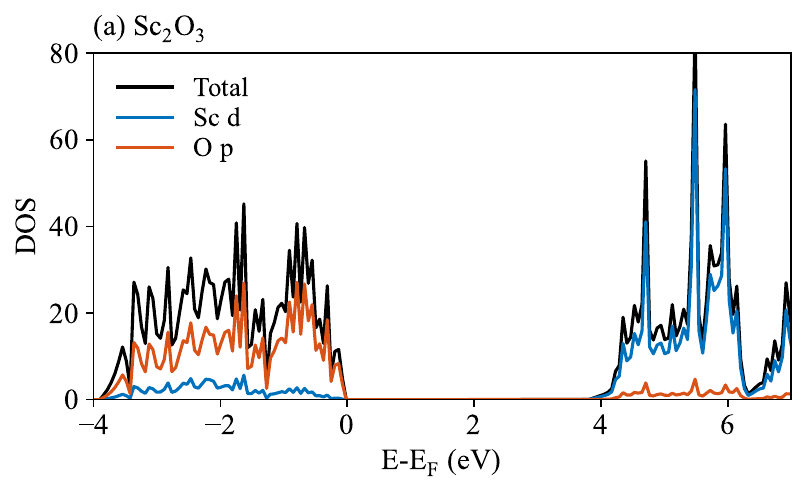}
    \includegraphics[width=0.51\linewidth]{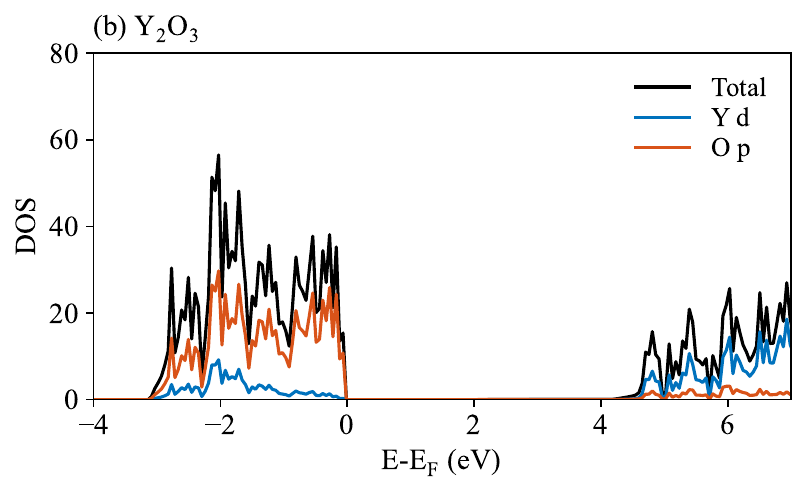}
    \includegraphics[width=0.51\linewidth]{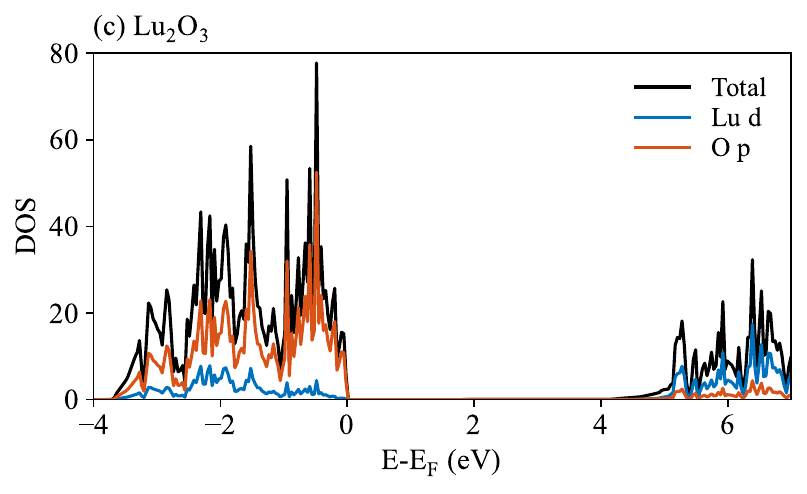}
    \caption{Calculated electronic density of states for the three bixbyite sesquioxides.}
    \label{fig:SMdos}
\end{figure}

\begin{table}[h]
\caption{\label{tab:structural}%
Calculated lattice parameters, band gaps, and surface energies 
$E_\mathrm{surf}$ for the three bixbyite sesquioxides. The (100) and (110) surface slab calculations applied $\Gamma$-centered 4$\times$4$\times$4 $k$-point mesh for Brillouin zone sampling, while only $\Gamma$ point is utilized for (111) surface.
}
\begin{ruledtabular}
\begin{tabular}{@{}cccccc}
\textrm{Material} &
{Lattice ($\mathrm{\AA}$)} &
{Band gap (eV)} &
\multicolumn{3}{c}{$E_\mathrm{surf}$ (J $\cdot$ m$^{-2}$)} \\
\cmidrule(lr){4-6}
& & & {(100)} & {(110)} & {(110)} \\[0.2em]
\colrule\\[-0.8em]
Sc$_2$O$_3$ & 9.9127 & 3.8295 (direct)   & 2.088 & 1.194 & 0.922 \\[0.2em]
Y$_2$O$_3$  & 10.6552 & 4.1176 (indirect) & 1.995 & 1.142 & 0.902 \\[0.2em]
Lu$_2$O$_3$ & 10.3527 & 4.0215 (indirect) & 2.288 & 1.290 & 1.022
\end{tabular}
\end{ruledtabular}
\end{table}

\bigskip

\begin{figure}[h]
    \centering
    \includegraphics[width=0.45\linewidth]{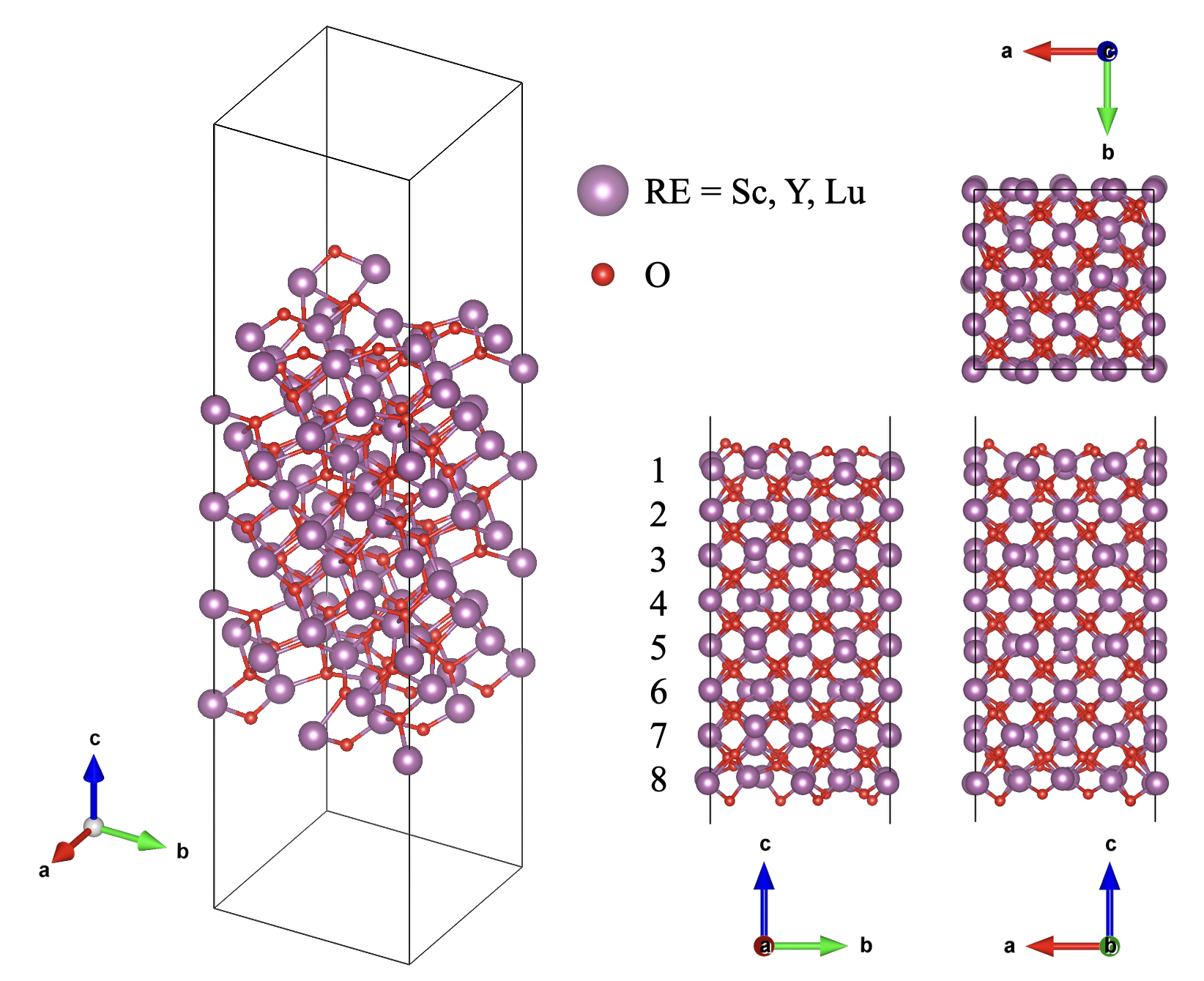} 
    \includegraphics[width=0.495\linewidth]{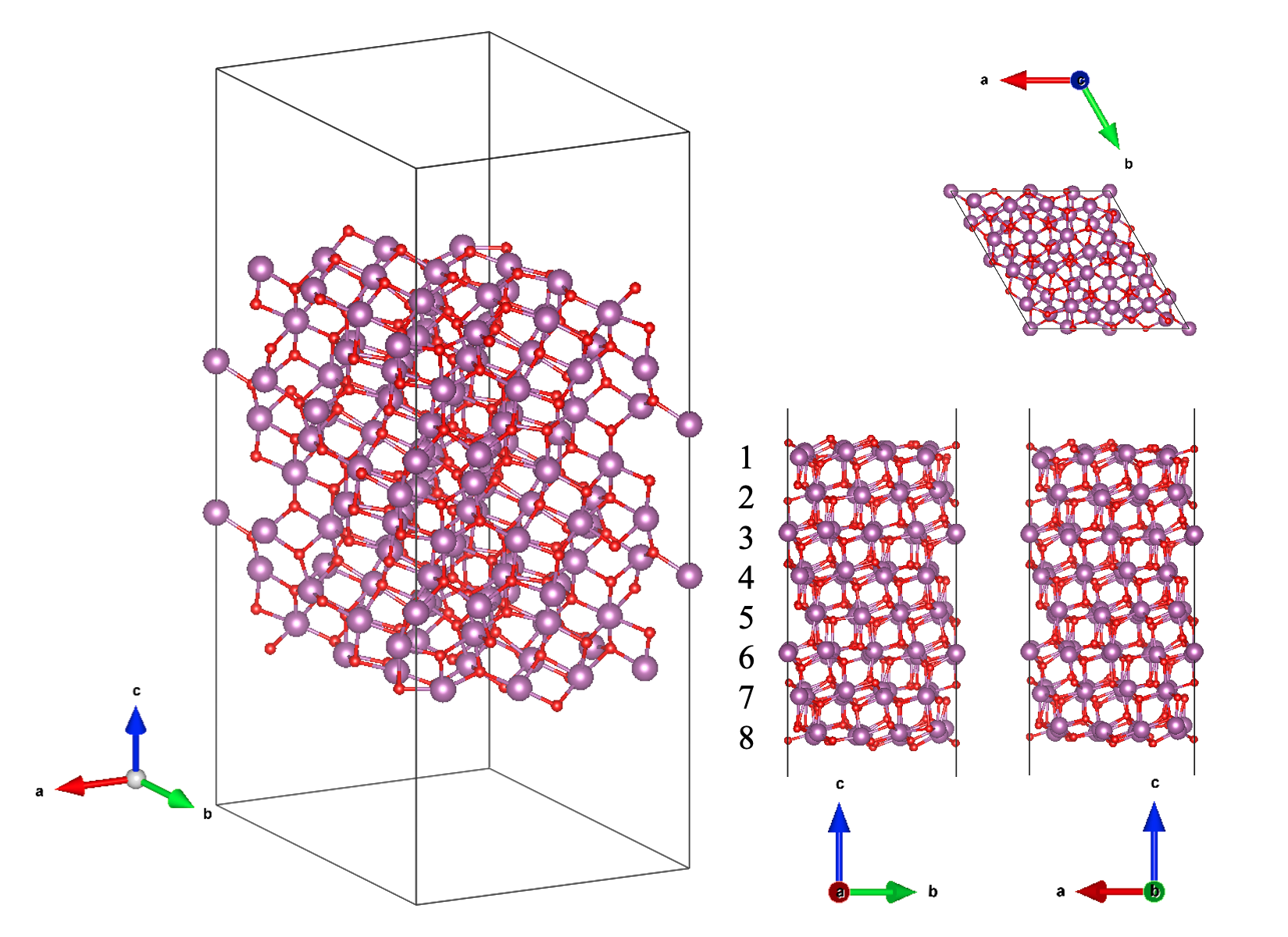} 
    \caption{Slab model of the (100) surface (left) and (111) surface (right) of \reto ~(RE = Sc, Y, Lu). A slab of 8 RE layers is used with a 10~\AA\ vacuum layer applied on both sides. Side views along the three crystallographic axes are also shown.}
    \label{fig:slab_100}
\end{figure}

\bigskip

\begin{table}[t]
\caption{\label{tab:statecriteria}%
Geometric criteria used to assign reaction states from AIMD trajectories.
$d(\text{O}_w\text{--RE}_s)$ denotes the distance between the water oxygen and nearest surface 
rare-earth atom; $d(\text{H}_w\text{--O}_w)$ the maximum O--H bond length within the water 
molecule; and $d(\text{H}_w\text{--O}_s)$ the maximum distance from each water hydrogen to the nearest 
surface oxygen.
}
\begin{ruledtabular}
\begin{tabular}{cl}
\textrm{State} & \textrm{Criteria} \\
\colrule\\[-0.8em]
IS &
$d(\text{O}_w\text{--RE}_s) > 3.0$~\AA \\[0.4em]
MS &
$d(\text{O}_w\text{--RE}_s) < 2.5$~\AA\
and $d(\text{H}_w\text{--O}_w) < 1.2$~\AA \\[0.4em]
DS &
$d(\text{O}_w\text{--RE}_s) < 2.5$~\AA\
and $d(\text{H}_w\text{--O}_w) > 1.3$~\AA\
and $d(\text{H}_w\text{--O}_s) > 2.0$~\AA \\[0.4em]
CS &
$d(\text{O}_w\text{--RE}_s) < 2.5$~\AA\
and $d(\text{H}_w\text{--O}_w) > 1.3$~\AA\
and $d(\text{H}_w\text{--O}_s) < 1.8$~\AA
\end{tabular}
\end{ruledtabular}
\end{table}

%\bibliographystyle{apsrev4-1} 
%\bibliography{reference}